\documentclass[%
aps,
twocolumn]{revtex4-2}
\usepackage{graphicx}
\usepackage{dcolumn}
\usepackage{bm}
\usepackage{amsmath,amssymb,amscd}         
\usepackage{dsfont}
\usepackage{hyperref}
\hypersetup{
	colorlinks=true,
	citecolor=blue,
	linkcolor=blue,
	urlcolor=magenta,
}
\newcommand{\nn}{\nonumber \\}
\newcommand{\nl}{\hspace{15pt}}
\usepackage{adjustbox}
\usepackage{tikz}

\def\diah{0.03}
\def\ii{\text{i}}

\def\ddt{{\text{d}\over \text{d}t}}
\def\ddtt{{\text{d}^2\over \text{d}t^2}}
\def\ed{\ .}

\def\co{\ ,}
\def\ns{\hspace{10pt}}

\def\eref#1{(\ref{#1})}

\def\Label#1{\label{#1}%
	\smash{\hbox to0pt{\raise1ex\hbox{\tiny[#1]}\hss}}}
 
\begin{document}
	
	
	\title{Noise effects on the diagnostics of quantum chaos}

	\author{Tingfei Li}
	\email{tfli@zju.edu.cn}
	\affiliation{%
		Zhejiang Institute of Modern Physics, Zhejiang University, Hangzhou, 310027, P. R. China
	}%
	\affiliation{Kavli Institute for Theoretical Sciences (KITS), University of Chinese Academy of Sciences, Beijing 100190, China}

	\date{\today}
	
	\begin{abstract}
		This paper investigates the effects of noise on the diagnostics of quantum chaos, focusing on three primary tools: the spectral form factor (SFF), Krylov complexity, and out-of-time correlators (OTOCs). Utilizing a closed quantum system model with white noise, we demonstrate that increasing noise strength leads to an exponential suppression of these diagnostic measures. Specifically, our findings reveal that in the strong noise limit, the SFF, two-point correlation function, and OTOCs become ineffective in distinguishing chaotic behavior. The SFF is particularly impacted, exhibiting a significant decay that obscures its ability to identify quantum chaos. This study highlights the challenges posed by environmental noise in accurately diagnosing quantum chaotic systems and suggests that traditional methods may require adaptation to remain effective in realistic open quantum systems. Our results underscore the need for further research into robust diagnostic techniques that can account for noise-induced effects in quantum chaotic systems.
	\end{abstract}
	
	\keywords{Quantum Chaos,Noise}
	\maketitle
	
	
	\section{Introduction}
	\paragraph{Motivation}
	Quantum chaos has emerged as a pivotal area of research, bridging the realms of quantum mechanics and classical chaotic systems. As a quantum analogue to classical chaos, it plays a critical role in understanding complex many-body quantum systems, and it has garnered significant attention due to its profound implications in fields such as black hole physics and quantum gravity. In classical systems, chaos is characterized by the exponential growth of uncertainty stemming from minor variations in initial conditions \cite{1963JAtS...20..130L,1971CMaPh..20..167R,Pesin_1977}, which can be easily conceptualized through the dynamics governed by ordinary differential equations (ODEs) \cite{Katok_Hasselblatt_1995}.
	In contrast, quantum systems evolve according to linear equations that govern the density matrix. This means that irregular motion in quantum mechanics cannot be characterized by extreme sensitivity to small changes in initial conditions \cite{Gutzwiller1990ChaosIC,qchaos2018}. This intrinsic property makes it difficult to identify chaotic behavior in quantum contexts, especially in open quantum systems, where certain density matrices may exhibit exponential decay. As a result, the observable differences between quantum states diminish over time, complicating the interpretation of quantum chaos as an emergent feature arising from slight initial perturbations.
	
	Due to these challenges, researchers have focused more on the statistical characteristics of quantum systems. Following the well-known conjecture by Bohigas, Giannoni, and Schmit \cite{PhysRevLett.52.1}, certain universal features of spectral fluctuations in classically chaotic systems have been found to align with predictions from random-matrix theory \cite{mehta1991random,Porter1965StatisticalTO}. Consequently, the spectral form factor (SFF) \cite{PhysRevLett.56.2449,PhysRevLett.67.1185,PhysRevLett.70.572,doi:10.1143/JPSJ.64.4059} has been proposed as a diagnostic tool for quantum chaos. In a broad sense, a chaotic system acts as an efficient information scrambler \cite{Stanford_2014, Maldacena_2016,Roberts_2015}. This is indicated by the exponential decay of out-of-time-order correlators (OTOCs) \cite{kitaev_talk,Shenker_2014,Roberts_2017,otoc2018,Harrow_2021}, efficient operator spreading \cite{Nahum_2018,Khemani_2018}, small fluctuations in purity \cite{Oliviero_2021}, and information scrambling \cite{Ding:2016txk,Hosur_2016}.

	In this paper, we examine three methods for diagnosing quantum chaos: OTOC, Krylov complexity~\cite{Parker_2019,Xu:2019lhc,Caputa:2021sib,Balasubramanian:2022tpr}, and SFF. The OTOC is a four-point correlator, while Krylov complexity is derived from a two-point function using a thermal state with inverse temperature $\beta$, emphasizing the importance of the energy spectrum's statistical behavior. The SFF, on the other hand, relies solely on the energy spectrum of the system.  Notably, the SFF remains a valuable indicator of quantum chaos even in open systems affected by energy dephasing \cite{Cornelius_2022}.
	
	This study investigates how noise affects these diagnostic measures by using a closed quantum system model exposed to white noise. Our results show that as the noise strength increases, there is an exponential suppression of the SFF, two-point correlation functions, and OTOCs, which ultimately makes them ineffective for distinguishing chaotic behavior in high noise conditions. This research highlights the need for more robust diagnostic techniques that can address the challenges posed by environmental noise in realistic open quantum systems, paving the way for future studies on quantum chaos in noisy environments.

	\paragraph{The Model}
	We conduct the calculations in the energy basis, ensuring that the Hamiltonian is diagonal. Our focus is primarily on noise without time correlation, specifically white noise. For simplicity, we consider a closed quantum system with a $D$-dimensional Hilbert space and examine the effects of noise. In this context, the time evolution is governed by the noisy Hamiltonian:
	\begin{align}
		H_{ij} = E_{i} \delta_{ij} + \eta_{ij}(t),i,j=1,2,\ldots,D
	\end{align}
	where $\eta_{ij}(t)$ is the white noise with zero mean and a non-vanishing variance
	\begin{align}
		\mathbb{E}(\eta_{ij}(t)) =0,~\mathbb{E}(\eta_{ij}(t) \eta_{kl}(t')) = \lambda_{ijkl} \delta(t-t')\ed
	\end{align}
	To distinguish from the state expectation \(\langle \bullet \rangle = \text{Tr}(\hat{\rho}(t)\bullet)\), we use \(\mathbb{E}(\bullet)\) to denote the ensemble average of the noise. 
	
	For general \(\lambda_{ijkl}\), obtaining an analytical solution is challenging. Therefore, in this paper, we focus on two special cases: \(\lambda_{ijkl} = \lambda_{ij} \delta_{il} \delta_{jk}\) (GUE noise) and \(\lambda_{ijkl} = \lambda_{ij} \frac{\delta_{ik} \delta_{jl} + \delta_{il} \delta_{jk}}{2}\) (GOE noise). We consistently use \(J\) to denote the strength of the noise, \textit{i}.\textit{e}., \(\lambda_{ijkl} \sim J\). Consequently, we sometimes denote the noise ensemble average of \(A\) as \(A_J \equiv \mathbb{E}(A)\). We adopt similar notation as in \cite{Tang:2024kpv}. 
	
	For generic \(k\)-replica observables in our model, we ultimately obtain an effective time evolution on \(2k\)-contours:
	
	\begin{align}
		\mathcal{U}_k\equiv \mathbb{E}[U_t^{\otimes k}\otimes U_t^{*\otimes k}]\equiv e^{\mathcal{L}_k t}
	\end{align}
	where \( U_t = \mathcal{T} e^{-\ii \int_0^t H(\tau) d\tau} \) is the noisy time evolution operator. Writing \( H = H_0 + \eta \), we denote \( U_t^{(0)} = e^{-\ii H_0 t} \) as the noise-free time evolution operator. For an operator \( A \), we define 
	\begin{align}
		A_t\equiv U_t^\dagger A U_t, A_t^{(0)}\equiv U_t^{(0)\dagger} A U_t^{(0)}\ed
	\end{align}
	In this paper, we primarily study the cases \( k = 1, 2 \), which are needed for the calculation of quantum chaos diagnostics.
	\paragraph{Summary of results}
	We investigate the effects of noise on three quantum chaos diagnostics: SFF, Krylov complexity, and OTOC. As an example, we consider GUE noise with \(\lambda_{ij} = \frac{J}{D}\). We find that the noise-averaged SFF consists of a spectral-dependent term and a universal noise term~(Eq.\eref{eq:GUE-sff} in main text)
	\begin{align}\label{eq:results-SFF}
		K_J(t)= e^{-Jt}K_{J=0}(t)+\frac{1}{D^2}\left(1-e^{-Jt}\right)
	\end{align}
	where \( K_{J=0}(t) \) is the SFF without noise. It is exponentially suppressed by the noise, indicating that the SFF is not an effective diagnostic for quantum chaos when the noise strength is sufficiently large. Meanwhile, the noise-averaged two-point function exhibits similar behavior (Eq.\eref{eq:2pt_GUE} in main text)
	\begin{align}\label{eq:results-2pt}
		C_J(t)=e^{-Jt}C_{J=0}(t)+{1\over D^2}\left(1-e^{-Jt}\right)\text{Tr}O\text{Tr}O^\dagger\ed
	\end{align}
	Thus, one can observe that noise induces oscillations in the Lanczos coefficients \( b_n \). When \( J \) is sufficiently large, the linear growth of \( b_n \) in the original system is no longer apparent. As for the two-replica dynamics, we find that
	\begin{align}\label{eq:results-U2}
		\left(\mathcal{U}_2(t)\right)_{ijkl;i'j'k'l'}=e^{-\ii E_{ijkl }t}\sum_{i=1}^{8}f_a(t)\mathsf{F}_a
	\end{align}
	where the factor \( e^{-\ii E_{ijkl} t} \) encodes information about the original system, while the noise effects are represented by the eight coefficients \( f_a(t) \) and the corresponding graphs \( \mathsf{F}_a \). With the explicit expression of \( \mathcal{U}_2 \), one can study OTOCs and other two-replica observables. For two traceless Hermitian operators \( A \) and \( B \), we find a simple formula for the disorder-averaged OTOC at infinite temperature (Eq.\eref{eq:otoc_GUE} in main text)
	\begin{align}\label{eq:results-OTOC}
		\text{OTOC}_{J}=e^{-2Jt}\text{OTOC}_{J=0}+O(1/D)\ed
	\end{align}
	Comparing Eq.~\eref{eq:results-SFF}, Eq.~\eref{eq:results-2pt}, and Eq.~\eref{eq:results-OTOC}, one finds that the noise introduces an exponential suppression factor, which can obscure the features of the original system for large \( Jt \). This may represent a universal effect of any noise. 
	\paragraph{Structure of the paper}
	We discuss \( k=1 \) in Section \eref{sec:GUE-noise} for GUE noise and Section \eref{sec:GOE-noise} for GOE noise. As a warm-up, we first study the constant case \( \lambda_{ij} = \frac{J}{D} \) in Subsection \eref{sec:GUE-const}, then consider the general case in Subsection \eref{sec:GUE-general}. Corresponding discussions for GOE noise are provided in Subsections \eref{sec:GOE-const} and \eref{sec:GOE-general}. Next, we address \( k=2 \) in Section \eref{sec:U2}, where, in addition to OTOC, we study the fluctuations of the SFF. Finally, we offer some discussion for future research in Section \eref{sec:conclusion}.

	\section{GUE noise}
	\label{sec:GUE-noise}
	Since we primarily consider the ordinary closed system, we require the noisy Hamiltonian to be Hermitian, \( H^\dagger = H \), at each time. Thus, we impose
	\begin{align}
		\eta_{ij}(t)=\eta_{ji}^*(t) \in \mathbb{C}\ed
	\end{align}
	For simplicity, we assume that all matrix elements of the noise term are independent Brownian Gaussian random variables with a vanishing mean and variances given by
	\begin{align}
		\mathbb{E}\left(\eta_{ij}(t)\eta_{kl}(t')\right) =\lambda_{ij} \delta_{il}\delta_{jk}\delta(t-t'),\lambda_{ij}=\lambda_{ji}>0\ed \label{eq:GUE-variance}
	\end{align}
	If we set \( \lambda_{ij} = \frac{1}{D} \) and \( E_i = 0 \) for all \( i \), we arrive at the Brownian Gaussian matrix ensembles studied in \cite{Tang:2024kpv}. Similar models have been discussed in \cite{guo2024complexityrandomness}. In this paper, we adopt the same notations as in \cite{Tang:2024kpv}. For generic \( k \)-replica observables in our model, we ultimately obtain an effective time evolution on \( 2k \)-contours:
	\begin{align}
		\mathcal{U}_k\equiv \mathbb{E}[U_t^{\otimes k}\otimes U_t^{*\otimes k}]\equiv e^{\mathcal{L}_k t}\ed
	\end{align}
	Next, we consider taking discrete time slices \( t_n = n \Delta t \), where \( \Delta t = \frac{t}{N} \) and \( n = 0, 1, 2, \ldots, N-1 \). At each time step, we have a random noise term \( \eta_{ij}(n \Delta t) \). For two times \( t_1 = n_1 \Delta t \) and \( t_2 = n_2 \Delta t \), we choose to regularize the delta function as follows:
	\begin{align}
		\delta(t_1-t_2)=\delta(n_1\Delta t - n_2 \Delta t) = \delta_{n_1n_2}{1\over \Delta t}\ed
	\end{align}
	So that Eq.\eref{eq:GUE-variance} becomes 
	\begin{align}
		\mathbb{E}\left(\eta_{ij}(n \Delta t)\eta_{kl}(m \Delta t)\right) ={1\over \Delta t}\lambda_{ij} \delta_{il}\delta_{jk}\delta_{nm}\ed \label{eq:regGUE-variance}
	\end{align}
    One can treat $\mathcal{U}_k$ as a $D^{2k}$-dimensional matrix generated by the operator $\mathcal{L}_k$, with the row and column indices labeled explicitly as $I = \{i_1, i_2, i_3, \ldots, i_{2k}\}$ and $J = \{j_1, j_2, j_3, \ldots, j_{2k}\}$
	\begin{align}
		\left[\mathcal{U}_k\right]_{IJ}=U_{i_1j_1}U^*_{i_2,j_2}\cdots U_{i_{2k-1}j_{2k-1}}U^*_{i_{2k},j_{2k}}
	\end{align}
    where we take the order of $\{i\}$ and $\{j\}$ on the right-hand side for convenience. Then, the equation of motion of $\mathcal{U}_k$ is simply 
	\begin{align}
		\partial_t \mathcal{U}_k(t) = \mathcal{L}_k. \mathcal{U}_k(t)\ed \label{eq:motioneq}
	\end{align}
	To illustrate the calculation, we first consider the simplest case of \(k=1\): 
	\begin{align}
		\mathcal{U}_1(t):=\mathbb{E}\left[U(t)\otimes U^*(t)\right], \mathcal{U}_{1;ij;i'j'}=\left[\exp{(\mathcal{L}_1 t)} \right]_{ij;i'j'}\ed 
	\end{align}
	One can obtain the expression for \(\mathcal{L}_1\) through direct calculation. From the definition, we have
		\begin{align}
			&U_{ij}(t+\Delta t )U_{kl}^*(t+\Delta t)\nn
			=& \sum_{u,v=1}^D\left(e^{-\ii H(t)\Delta t} \right)_{iu}\left(e^{\ii H^*(t)\Delta t} \right)_{kv}U_{uj}(t)U_{vl}^*(t)\ed 
		\end{align}
	Next, we employ the Taylor series expansion and take the noise ensemble average on both sides, retaining terms up to linear order in \(\Delta t\). Thus, we obtain
	\begin{widetext}
			\begin{equation}
				\begin{aligned} 
					&\nl \mathbb{E}\left[U_{ij}(t+\Delta t)U_{kl}^{*}(t+\Delta t)\right]\\
					& =\mathbb{E}\left[\sum_{u,v=1}^{D}\left(e^{-\ii H(t)\Delta t}\right)_{iu}\left(e^{\ii H^{*}(t)\Delta t}\right)_{kv}U_{uj}(t)U_{vl}^{*}(t)\right]\\
					& =\mathbb{E}\Bigg[\sum_{u,v=1}^{D}\left(\mathbb{I}-\ii H(t)\Delta t-\frac{1}{2}H^{2}(t)\Delta t^{2}\right)_{iu}\left(\mathbb{I}+\ii H^{*}(t)\Delta t-\frac{1}{2}H^{*2}(t)\Delta t^{2}\right)_{kv}U_{uj}(t)U_{vl}^{*}(t)\Bigg]\\
					& =\sum_{u,v=1}^{D}\mathbb{E}\bigg[\left(\delta_{iu}-\ii E_{i}\delta_{iu}\Delta t-\ii\eta_{iu}(t)\Delta t\right)\left(\delta_{kv}+\ii E_{k}\delta_{kv}\Delta t+\ii\eta_{kv}^{*}(t)\Delta t\right)\bigg]\mathbb{E}\left[U_{uj}(t)U_{vl}^{*}(t)\right]\\
					& =\sum_{u,v=1}^{D}\bigg[\delta_{iu}\delta_{kv}-\ii E_{i}\delta_{iu}\delta_{kv}\Delta t+\ii E_{k}\delta_{iu}\delta_{kv}\Delta t-\frac{1}{2}\sum_{s}\lambda_{is}\delta_{iu}\delta_{kv}\Delta t-\frac{1}{2}\sum_{s}\lambda_{ks}\delta_{iu}\delta_{kv}\Delta t+\lambda_{ik}\delta_{ik}\delta_{uv}\Delta t\bigg]\times\mathbb{E}\left[U_{uj}(t)U_{vl}^{*}(t)\right]\nonumber
				\end{aligned}
		\end{equation}
	\end{widetext}
	where \(\mathbb{I}\) is the identity matrix. We have utilized the fact that white noise has no time correlation, such that \(\mathbb{E}(A(t)B(t')) = \mathbb{E}(A(t))\mathbb{E}(B(t'))\) for \(t \neq t'\). By comparing the result with the differential equation given in Eq.\,\eqref{eq:motioneq} and recognizing \(\mathcal{L}_1\), we obtain
	\begin{align}
		\mathcal{L}_{1;ij;i'j'}=w_{ij}\delta_{ii'}\delta_{jj'}+\lambda_{ii'}\delta_{ij}\delta_{i'j'}\ed \label{eq:L1}
	\end{align}
	Here $w_{ij}\equiv -\ii E_{i}+\ii E_{j}-{1\over 2}\sum_{k=1}^D \left(\lambda_{ik}+\lambda_{jk}\right)$.
	For simplicity we use the graph representation 
	\begin{equation}
		\begin{aligned}
			\mathcal{L}_{1;ij;i'j'}=w_{ij}\adjincludegraphics[valign=c, height=\diah\textwidth]{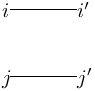}+\lambda_{ii'}\adjincludegraphics[valign=c, height=\diah\textwidth]{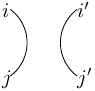}
		\end{aligned}
		\label{eq:L1exp}
	\end{equation}
	where each line with two indices \(i,j\) at the endpoints of the graph represents a ``propagator" \textit{i}.\textit{e},  $\delta_{ij}$. Similarly, one can derive expressions for other operators \(\mathcal{L}_k\) with \(k \geq 1\). One may regard \(\mathcal{L}_k\) as the effective time-independent Hamiltonian of an auxiliary system, while \(\mathcal{U}_k\) represents the imaginary time evolution. In principle, to obtain the dynamics, one needs to diagonalize \(\mathcal{L}_k\).
	
	The task becomes significantly more challenging when the expression for \(\mathcal{L}_k\) becomes too complex for large \(k\). For \(k=1\) and \(k=2\), however, it is promising to find explicit expressions for \(\mathcal{U}_k\).
	\subsection{$\lambda_{ij}=\text{const}$}
	\label{sec:GUE-const}
	To illustrate the key process of the calculation clearly,  we first consider the simplest case $\lambda_{ij}= {J\over D}$, so that Eq.\eref{eq:L1} becomes
	\begin{align}
		\mathcal{L}_{1;ij;i'j'}=w_{ij}\delta_{ii'}\delta_{jj'}+{J\over D}\delta_{ij}\delta_{i'j'}, w_{ij}=-\ii E_{i}+\ii E_{j}-J\ed
	\end{align}
	We consider evaluating \(\mathcal{U}_1\) by using recursion relations. Firstly, we make an ansatz:
	\begin{align}
		\left(\mathcal{L}^n_{1}\right)_{ij;i'j'}=c_{n}^{(ij)}\delta_{ii'}\delta_{jj'}+d_{n}^{(ii')}\delta_{ij}\delta_{i'j'}\ed
	\end{align}
	It is more convenient to use the graph representation
	\begin{equation}
		\begin{aligned}
			\left(\mathcal{L}^n_{1}\right)_{ij;i'j'}=c_n^{(ij)}\adjincludegraphics[valign=c, height=\diah\textwidth]{id2.pdf}+d_n^{(ii')}\adjincludegraphics[valign=c, height=\diah\textwidth]{cross2.pdf}\ed
		\end{aligned}
		\label{eq:L1set}
	\end{equation} 
	To obtain the recursion relation, we multiply the equation with $\mathcal{L}_1$, we find the results for $\left(\mathcal{L}^{n+1}_{1}\right)_{ij;i'j'}$
	\begin{equation}
		\begin{aligned}
			&\left(\mathcal{L}^{n+1}_{1}\right)_{ij;i'j'}\nn
			=&w_{ij}c_n^{(ij)}\adjincludegraphics[valign=c, height=0.03\textwidth]{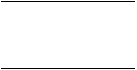}+w_{ii}d_n^{(ii')}\adjincludegraphics[valign=c, height=0.03\textwidth]{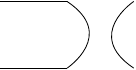} \\
			& + {J\over D}c_n^{(i'i')}\adjincludegraphics[valign=c, height=0.03\textwidth]{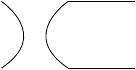}+ {J\over D}\sum_k d^{(ki')}
			\adjincludegraphics[valign=c, height=0.03\textwidth]{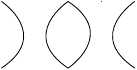} \\
			=&w_{ij}c_n^{(ij)}\adjincludegraphics[valign=c, height=\diah\textwidth]{id2.pdf}+\left[{J\over D}c_{n}^{(ii)}+{J\over D}\sum_{k}d_{n}^{(ki')}-{J}d_{n}^{(ii')}\right]\adjincludegraphics[valign=c, height=\diah\textwidth]{cross2.pdf}
		\end{aligned}
		\label{eq:L1recur}
	\end{equation} 
	where we have used \(w_{ii} = -J\). Comparing the result with the ansatz given in Eq.\,\eqref{eq:L1set}, we obtain the recursion relations
	\begin{align}
		c_{n+1}^{(ij)}&=w_{ij}c_n^{(ij)}, d_{n+1}^{(ii')}=-{J}d_{n}^{(ii')}+{J\over D}c_{n}^{(ii)}+{J\over D}\sum_{k}d_{n}^{(ki')}\ed\nonumber
	\end{align}
	With the initial condition \(c_{1}^{(ij)} = w_{ij}\), we easily find \(c_n^{(ij)} = w_{ij}^n\). Thus, the recursion relation for \(d_{n}^{(ii')}\) becomes
	\begin{align}
		d_{n+1}^{(ii')}=-{J}d_{n}^{(ii')}+{J\over D}\left(-J\right)^n+{J\over D}\sum_{k}d_{n}^{(ki')}\ed
	\end{align}
    Using initial condition $d_1^{(ii')}={J\over D}$ one can find $d_n^{(ii')}$ is independent of its indices, so we can denote $d_n = d_n^{(ii')}$ and obtain 
	\begin{align}
		d_{n+1}={J\over D}\left(-J\right)^{n}\ed
	\end{align}
	Having established \(c_n^{(ij)}\) and \(d_n^{(ii')}\), it is straightforward to construct \(\mathcal{U}_1\)
	\begin{equation}
		\begin{aligned}
			&\mathcal{U}_{1;ij;i'j'}(t)\equiv \sum_{n=0}^{\infty }{\left(\mathcal{L}_1^n\right)_{ij;i'j'}t^n \over n!}\nn
			=&\delta_{ii'}\delta_{jj'}+\sum_{n=1}^\infty\frac{\left(w_{ij}\right)^{n}\delta_{ii'}\delta_{jj'}+{J\over D}\left(-J\right)^{n-1}\delta_{ij}\delta_{i'j'}}{n!}t^{n}\\
			=&\exp\left(w_{ij}t\right)\adjincludegraphics[valign=c, height=\diah\textwidth]{id2.pdf}+\frac{1}{D}\left[1-\exp\left(-Jt\right)\right]\adjincludegraphics[valign=c, height=\diah\textwidth]{cross2.pdf}\ed
		\end{aligned}
		\label{eq:U1-exp}
	\end{equation}
	With the explicit expression of \(\mathcal{U}_1\), one can study the effects of noise on the SFF and Krylov complexity, and other noise effects. 
	\begin{figure}[h]
		\begin{center}
			\includegraphics[width=0.43\textwidth]{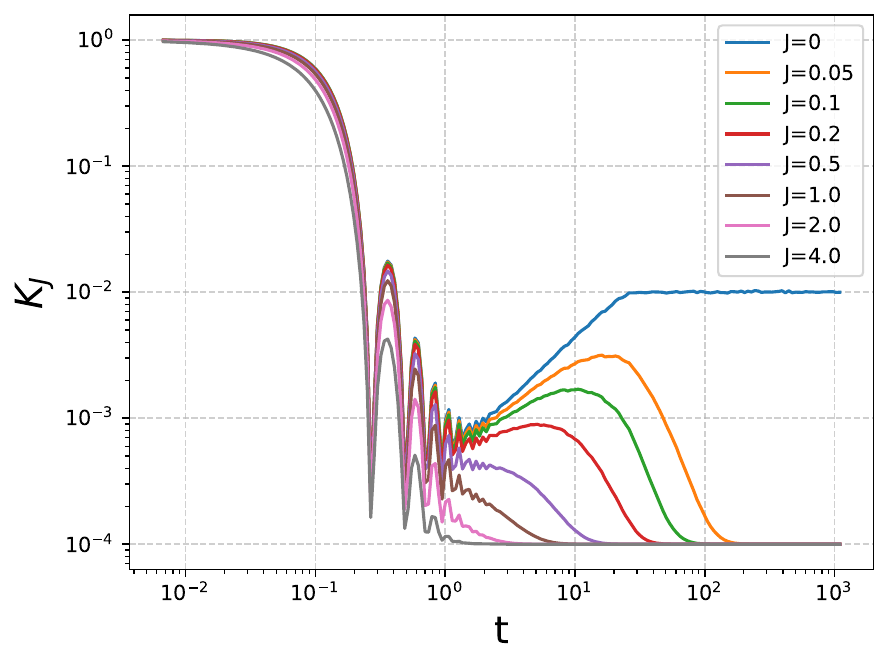}
			\caption{The effect of noise on SFF is studied, where the spectrum \(\{E_i\}_{i=1}^{D=100}\) is drawn from the GUE. We perform an ensemble average over \(8000\) realizations and use Eq.\eref{eq:GUE-sff} to generate the plot. It is evident that strong noise obscures the characteristic linear ramp of the SFF and the late-time plateau altitude falls to \(\frac{1}{D^2}=10^{-4}\) for non-vanishing $J$.}
			\label{fig:SFFJ}
		\end{center}
	\end{figure}
    \begin{figure}[h]
    	\begin{center}
    		\includegraphics[width=0.43\textwidth]{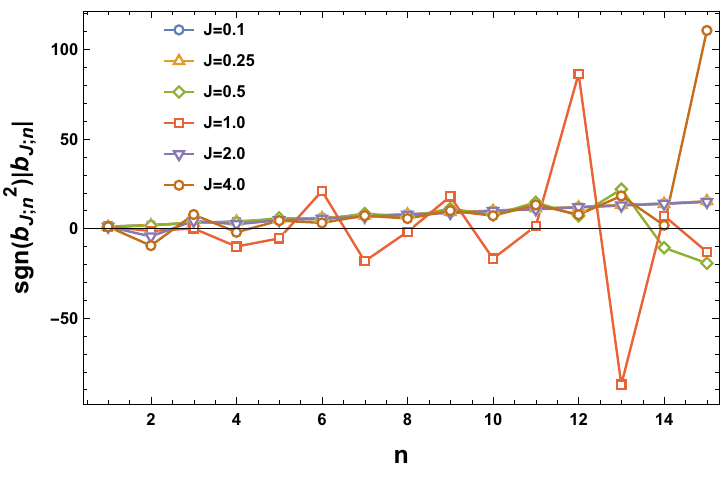}
    		\caption{The effect of noise on the \textit{signed} Lanczos coefficients \(\text{sgn}(b_{n;J}^2)|b_{n;J}|\) is examined for a simple model with \(C(t) = \text{sech}(\alpha t)\), where \(\alpha = 1\) and \(\text{Tr} O \text{Tr} O^\dagger = D^2\). We find that \(b_{n;J}\) is either real (\(\text{sgn}(b_{n;J}^2) = 1\)) or purely imaginary (\(\text{sgn}(b_{n;J}^2) = -1\)).}
    		\label{fig:bnJ}
    	\end{center}
    \end{figure}
	\paragraph{SFF}
	The noise-averaged SFF is defined as 
	\[
	K_J(t) \equiv \frac{1}{D^2} \mathbb{E}\left(\text{Tr}\, U(t) \, \text{Tr}\, U^\dagger(t)\right).
	\]
	This requires us to impose \(i = i'\) and \(j = j'\) in \(\mathcal{U}_{1;ij;i'j'}(t)\) and sum over \(i\) and \(j\). It is straightforward to obtain
	\begin{align}\label{eq:GUE-sff}
		K_J(t)= e^{-Jt}K_{J=0}(t)+\frac{1}{D^2}\left(1-e^{-Jt}\right)\co
	\end{align}
	where \(K_{J=0}(t)\) denotes the SFF without noise, and we have normalized \(K_J(t=0) = 1\). Generally speaking, \(K_{J=0}(t)\) will exponentially decay at early times and reach a plateau value of \(\frac{1}{D}\) for late times. A ramp occurring at the time scale \(t_{\text{ramp}} = \log(D)\) is regarded as a signal for quantum chaos. This picture dramatically changes when we introduce noise. Firstly, by taking the limit $t\to \infty$ in Eq.\eref{eq:GUE-sff}, one find the late-time plateau altitude falls to \(\frac{1}{D^2}\) for any \(J > 0\). Moreover, the ramp behavior in \(K_J(t)\) disappears for large \(J\), indicating that the SFF is not a good diagnostic for quantum chaos when the noise strength is sufficiently large. Taking the GUE as an example of the SFF with a ramp, we plot \(K_J\) in Fig.\ref{fig:SFFJ}. The ramp suppression also occurs in closed systems with non-Hermitian Hamiltonian deformation \cite{Matsoukas-Roubeas:2022odk} or open quantum systems governed by Lindblad dynamics \cite{PhysRevB.103.064309, Bhattacharyya_2023, PhysRevResearch.4.033093}.  
	\paragraph{Krylov complexity}
	Besides the SFF, another diagnostic tool, the Krylov complexity~\cite{Parker_2019} (Lanczos coefficients $b_n$), can be calculated from the disorder-averaged two-point function 
	\begin{align}
		C_{J}(t)\equiv \frac{1}{D}\mathbb{E}\left[\text{Tr}\left(O^{\dagger}O(t)\right)\right]\ed
	\end{align}
	Here for simplicity, we just consider case of the infinite temperature. Notice that 
	\begin{equation}
		\begin{aligned}
			&\mathbb{E}\left[\text{Tr}\left(A.B(t)\right)\right]=\mathbb{E}\left[\text{Tr}\left(AU^{\dagger}(t)B U(t)\right)\right]\nn
			&=\sum_{ijkl}^D A_{ij}B_{kl}\left(\mathcal{U}_1\right)_{li;kj}=\adjincludegraphics[valign=c, height=\diah\textwidth]{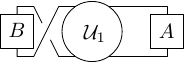}\co
		\end{aligned}
	\end{equation}
	where we have used the graph representation for an operator 
	\begin{align}
		A_{ij}=\adjincludegraphics[valign=c, height=0.04\textwidth]{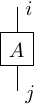}\ed
	\end{align}
	Using the expression in Eq.\eref{eq:U1-exp}, we have 
	\begin{align}\label{eq:2pt_GUE}
		C_J(t)=e^{-Jt}C_{J=0}(t)+{1\over D^2}\left(1-e^{-Jt}\right)\text{Tr}O\text{Tr}O^\dagger\ed
	\end{align}
	The moments $\mu_k$ are obtained by taking the Taylor series expansion at \(-it\)
	\begin{align}
		C_{J=0}(-it)=\sum_{n=0}^{\infty} \mu_{2n}{t^{2n}\over (2n)!}, C_{J}(-it)=\sum_{k=0}^{\infty} \mu_{J;k}{t^{k}\over k!}\ed
	\end{align}
	Thus, we find that the odd moments \(\mu_{J;2j+1}\) do not vanish when \(J \neq 0\). Generally, we have  
	\begin{align}
		\mu_{J;k}&=\sum_{i=0}^{k}\Bigg[\frac{k!(\sqrt{-1})^{i}}{i!(k-i)!}J^{i}\mu_{k-2i}\nn
		&\nl+\frac{\text{Tr}O\text{Tr}O^{\dagger}}{D^{2}}\left(\delta_{k,0}-\frac{(\sqrt{-1})^{k}J^{k}}{k!}\right)\Bigg]\ed
	\end{align}
	When \(J = 0\), we find that the Lanczos coefficients \(a_n = 0\). Thus, one can calculate \(b_n\) via the recursion relation below~\cite{Parker_2019}:
	\begin{equation}
		\begin{aligned}
			b_{n}&=\sqrt{M_{2n}^{(n)}}\co\\M_{2k}^{(m)}&=\frac{M_{2k}^{(m-1)}}{b_{m-1}^{2}}-\frac{M_{2k-2}^{(m-2)}}{b_{m-2}^{2}},k=m,\ldots,n\co\\M_{2k}^{(0)}&=\mu_{2k},b_{-1}=b_{0}:=1,M_{2k}^{(-1)}:=0\ed
		\end{aligned}
	\end{equation}
    It is suggested that the linear growth of \(b_n\) with \(n\) can be considered a signature of quantum chaos. If one applies the same method to the case with noise, we wonder what the outcome will be.  In the absence of noise, the Lanczos coefficients exhibit a linear growth \(b_n = n\). However, for large \(J\), the linear growth is disrupted, giving rise to oscillatory behavior as a function of \(n\). The oscillatory behavior is intricate, as shown in the Fig.\ref{fig:bnJ}: the data for \(J = 1.0\) displays much larger oscillations compared to the case of \(J = 2.0\). Strictly speaking, when \(J \neq 0\), we have \(a_n \neq 0\), which necessitates the use of the algorithm described in \cite{nandy2024quantumdynamicskrylovspace}. However, since a similar phenomenon can be observed in this case, we omit the explicit demonstration here.
	
	\paragraph{$r$-parameter}
	The $r$-parameter has been proposed as a novel signature for quantum chaos, as demonstrated in prior studies~\cite{Oganesyan_2007,PhysRevLett.110.084101}. For a system with ordered energy levels $\{e_n\}_{n=1}^D$, where $s_n = e_{n+1} - e_n$ represents the nearest-neighbor level spacing, we define the following quantities:
	\begin{align}
		r_n = \frac{s_n}{s_{n-1}}, \quad \widetilde{r}_n = \min\left(r_n, \frac{1}{r_n}\right).
	\end{align}
	To analyze the distribution of $r_n$ or $\widetilde{r}_n$, it is necessary to define the eigenvalues of the Hamiltonian $H$ for a noisy system. Given that the Hamiltonian entries in this context are Brownian random variables, we introduce a noise-averaged Hamiltonian as follows: Starting with $O(0) = \text{diag}(\{E_i\})$, we consider measure $O(t)$ after a finite time $t$. This allows us to define the effective Hamiltonian as
	\begin{align}\label{eq:Heff}
		H^{\text{eff}}(t) = \mathbb{E}\left(O(t)\right)\vert_{O(0)=\text{diag}(\{E_i\})},
	\end{align}
	where $\mathbb{E}$ denotes the expectation value over the noise ensemble. 
    It is easy to find 
    \begin{align}
    	H_{ij}^{\text{eff}}(0)=E_i\delta_{ij}\co
    \end{align}
    and for $t>0$
    \begin{align}\label{eq:Heff-fig}
    	(H^{\text{eff}}(t))_{ij}&=\adjincludegraphics[valign=c, height=\diah\textwidth]{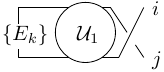}\nn
    	&=e^{-Jt}E_i\delta_{ij}+\overline{E} \left(1-e^{-Jt}\right)\delta_{ij}\ed
    \end{align}
    where $\overline{E}={\sum_{k=1}^D E_k \over D}$. So that the eigenvalues of $H^{\text{eff}}(t)$ are 
    \begin{align}\label{eq:newEi}
    	E_{J;i}=e^{-Jt}E_i+\overline{E}\left(1-e^{-Jt}\right)\ed
    \end{align}
    The noise-averaged level spacing is given by $s_{J;n} = e^{-Jt} s_n$, while $r_n$ remains invariant. This implies that $r_n$ continues to serve as an effective diagnostic tool for quantum chaos, provided that the effective Hamiltonian $H^{\text{eff}}_{ij}(t)$ can be experimentally measured. However, for large values of $J\Delta t \gg 1$, where $\Delta t$ represents the time cost of measurement, the noise-averaged level spacing $s_{J;n}$ becomes exponentially suppressed ($s_{J;n} \ll 1$). In this regime, accurately measuring $s_{J;n}$ becomes experimentally challenging, potentially compromising the validity of the $r$-parameter as a reliable diagnostic.
	\paragraph{Noise-induced ergodicity }
	One important effect of noise is that it enables the transition between different energy levels, which is forbidden for a closed system governed by pure Hermitian dynamics. In such a system, an energy eigenstate with eigenvalue \(E_i\) will never evolve into another energy state with eigenvalue \(E_f \neq E_i\). Once noise is considered, transitions between different states can occur, leading to diffusion in the Hilbert space.
	
	Consider the initial state to be \(|E_i\rangle\). We want to calculate the mean transfer probability to another state \(|E_j\rangle\); it is easy to find 
	\begin{align}
		\mathbb{E}(P_{j\leftarrow i})&= \mathbb{E}\left(U_{ji}(t)U_{ji}^* (t)\right)\nn
		&= \delta_{ij}\exp\left(-Jt\right)+\frac{1}{D}\left[1-\exp\left(-Jt\right)\right]\co
	\end{align}
    one can see it has no relation of the spectrum of the system for we consider the noise with infinite temperature. One can see that by taking $J>0,t\to \infty$, and find  $\mathbb{E}(P_{j\leftarrow i})\to {1\over D}$ lead to the canonical ensemble with infinite temperature. 
    And we can find the  variance of the transfer probability 
    \begin{align}
    	&\mathbb{E}\left(P_{j\leftarrow i}^{2}\right)-\mathbb{E}\left(P_{j\leftarrow i}\right){}^{2}\nn
    	&=\mathbb{E}\left(U_{ji}^{2}(t)U_{ji}^{*2}(t)\right)-\mathbb{E}\left(U_{ji}(t)U_{ji}^{*}(t)\right){}^{2}\nonumber
    \end{align}
   is also universal. 
    Thus, we may need to consider general noise. Imagine that the noise arises from collisions between the system and particles in the environment. We assume that the distribution of the environmental particles follows the Gibbs ensemble
    \begin{align}
    	\rho(E) \propto  e^{-\beta E}\ed
    \end{align}
    Here, \(E\) is the energy of the environmental particles, and \(\beta = \frac{1}{T}\) is the inverse temperature of the environment. Thus, the transition probability of the system from level \(E_i\) to \(E_j\) is proportional to \(\exp(-\beta |E_i - E_j|)\) for a short time. We can then set
    \begin{align}
    	\mathbb{E}\left( \eta_{ij}(t)\eta_{kl}(t')\right) ={J\over N}\exp(-\beta |E_i-E_j|) \delta_{il}\delta_{jk}\delta(t-t')\ed\nonumber
    \end{align}
    This means that it is harder to make a transition between states with a larger energy gap. The case we discussed in this section can be regarded as the infinite temperature limit \(\beta \to 0\). 
    
    In addition to the transition probability, we can calculate the mean return probability
	\begin{align}
		P_S(t) \equiv  {1\over D_S}\sum_{k=1}^{D_S}\text{Tr}\left(\Pi_k(t) \Pi_k\right)
	\end{align}
	where the $D_S$ projectors $\{\Pi_k\}$ form a complete decomposition of the whole Hilbert space via
	\begin{align}
		\sum_{k=1}^{D_S}\Pi_k= \mathbb{I}, \Pi_k\Pi_l = \delta_{kl}P_k, P_k=P_k^\dagger\ed 
	\end{align}
	For a closed system, it is proved that \(P_S(t)\) can be used as a bound for SFF \cite{Vikram_2024, vikram2024proofuniversalspeedlimit}
	\begin{align}
		P_S(t) \ge K(t)\ed
	\end{align}
	Now, we take the noise ensemble average on both sides to obtain
	\begin{align}
		P_{S;J}(t)\equiv \mathbb{E}\left(P_S(t)\right) \ge K_J(t)\equiv \mathbb{E}\left(K(t)\right)\ed
	\end{align}
	For example, we consider \(D_S = D\) and choose the eigenstate of the Hamiltonian. Then $\left(\Pi_k\right)_{ij} =\delta_{ik}\delta_{jk}$ 
	\begin{align}
		P_{S;J}(t)=\exp\left(-Jt\right)+\frac{1}{D}\left[1-\exp\left(-Jt\right)\right]\ed
	\end{align}
	One can see that for large time \(t_s \sim \frac{1}{J} \log \left( \frac{D}{2} \right)\), \(P(t) \to \frac{1}{D}\), indicating that the system is heated to infinite temperature.

	\subsection{General case}
	\label{sec:GUE-general}
	As discussed in the last section, it is necessary to consider more general noise with the invariance \(\lambda_{ij} \neq \text{const}\). Using the same ansatz for \(\mathcal{U}_1\), after some similar calculations, we have
	\begin{align}
		c_{n+1}^{(ij)}=w_{ij}c_n^{(ij)},d_{n+1}^{(ii')}&=w_{ii}d_{n}^{(ii')}+\lambda_{ii'}c_{n}^{(i'i')}+\sum_{k}\lambda_{ik}d_{n}^{(ki')}\ed \nonumber
	\end{align}
	Notice that the expression for \(w_{ij}\) is more complex
	\begin{align}
		w_{ij}\equiv -\ii E_{i}+\ii E_{j}-{1\over 2}   \left(J_i+J_j\right)
	\end{align}
	where we have defined \(J_i = \sum_{k=1}^D \lambda_{ik}\). Then, using the same ansatz for \(\mathcal{U}_1\), it is straightforward to find
	\begin{align}
		c_{n}^{(ij)}=w_{ij}^{n},c_{n}^{(ii)}=\left(-J_i\right)^{n}\ed 
	\end{align}
	For simplicity, we regard \(d_{n}^{(ii')}\) as a matrix with row index \(i\) and column index \(j\). Then, we have 
	\begin{align}
		d_{n+1}=P.d_{n}+Q_{n} \label{eq:dn-recur}
	\end{align}
	where $P,Q_n$ are two matrices 
	\begin{align}
		P_{ij}=-J_i\delta_{ij}+\lambda_{ij},\left(Q_{n}\right)_{ij}=\left(-J_j\right)^{n}\lambda_{ij}\ed
	\end{align}
	Solving the recursion relation Eq.\eref{eq:dn-recur}, we get
	\begin{align}
		d_n = \sum_{k=0}^{n-1}P^{n-1-k}.Q_{k}
	\end{align} 
	In the case of \(\lambda_{ij} = \frac{J}{D}\), we have \(J_j = J\) and \(P . Q_k = \mathbf{0}\), so the solution can be simplified to \(d_n = Q_{n-1}\). However, for the general case, there is no closed form for \(d_n\). As before, we can attempt to construct \(\mathcal{U}_1\)
	\begin{equation}
		\begin{aligned}
			\mathcal{U}_{1}=&\delta_{ii'}\delta_{jj'}+\sum_{n=1}^{\infty}\frac{w_{ij}^{n}\delta_{ii'}\delta_{jj'}+\sum_{k=0}^{n-1}P^{n-1-k}.Q_{k}\delta_{ij}\delta_{i'j'}}{n!}t^{n}\\=&\exp\left(w_{ij}t\right)\delta_{ii'}\delta_{jj'}+\sum_{n=1}^{\infty}\frac{\sum_{k=0}^{n-1}P^{n-1-k}.Q_{k}\delta_{ij}\delta_{i'j'}}{n!}t^{n}\ed \nonumber
		\end{aligned}
	\end{equation}
	Thus, one must deal with the exponent and the inverse of the matrix \(P\). In principle, this is a challenging task for large \(D\). In fact, it is simpler to obtain a closed form for \(\mathcal{U}_1(t)\) by solving its differential equation
	\begin{align}
		\partial_t \mathcal{U}_1(t)= \mathcal{L}_1.\mathcal{U}_1(t)\ed
	\end{align}
	Using the ansatz 
	\begin{align}
		\mathcal{U}_1=A_{ij}\adjincludegraphics[valign=c, height=\diah\textwidth]{id2.pdf}+B_{ii'}\adjincludegraphics[valign=c, height=\diah\textwidth]{cross2.pdf}\co
	\end{align}
    we then find
	\begin{align}
		\partial_{t}A_{ij}&=w_{ij}A_{ij}\co\nn
		\partial_{t}B_{ii'}&=w_{ii}B_{ii'}+\lambda_{ii'}A_{i'i'}+\sum_{s}\lambda_{is}B_{si'}\ed
	\end{align}
	Solving these equations, we have 
	\begin{align}
		A_{ij}&=\exp\left(w_{ij}t\right),B=\exp\left(Pt\right).C(t),\nn
		C(t)&=\int_{0}^{t}dt\exp\left(P^{-1}t\right).Q(t),~Q(t)_{ij}=\lambda_{ij}\exp(w_{jj}t)\ed\nonumber
	\end{align}
	Here we consider a simple case $J_j=J$, so we finally have 
	\begin{equation}
		\begin{aligned}
			K_{J}(t)&=e^{-Jt}K_{J=0}(t)\\
			&\nl +\frac{1}{D^{2}}\text{Tr}\left[\exp\left(Pt\right).\int_{0}^{t}dt\exp\left(P^{-1}t\right).Q(t)\right]\ed\nonumber
		\end{aligned}
	\end{equation}
	The formula is similar to that of the constant case. One can also evaluate the effect of noise on the two-point function.
	\section{GOE noise}
	\label{sec:GOE-noise}
	If the system we consider has other symmetries, such as time-reversal symmetry, we need to impose \(H = H^\dagger\) for \(\mathcal{T} = 1\), so that
	\begin{align}
		H_{ij}=E_i\delta_{ij}+\eta_{ij}(t),~\eta_{ij}(t)=\eta_{ji}(t) \in \mathbb{R}\co
	\end{align}
	where the variance of the noise is 
	\begin{align}
		\mathbb{E}\left(\eta_{ij}(t)\eta_{kl}(t')\right)=\lambda_{ij}\frac{\delta_{ik}\delta_{jl}+\delta_{il}\delta_{jk}}{2}\delta(t-t')\ed
	\end{align}
	After similar calculation for GUE case, we have 
	\begin{equation}
		\begin{aligned}
			\mathcal{L}_{1;ij;i'j'}=w_{ij}\adjincludegraphics[valign=c, height=\diah\textwidth]{id2.pdf}+{1\over 2}\lambda_{ii'}\adjincludegraphics[valign=c,height=\diah\textwidth]{cross2.pdf}+{1\over 2}\lambda_{ii'} \adjincludegraphics[valign=c, height=\diah\textwidth]{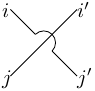}\ed\nonumber
		\end{aligned}
		\label{eq:GOE-L1}
	\end{equation}
	\subsection{$\lambda_{ij}=\text{const}$}
	\label{sec:GOE-const}
	We first consider the simplest case, where the noise is independent of the spectrum
	\begin{align}
		\lambda_{ij}={J\over D}\ed
	\end{align}
	For this case we have 
	\begin{align}
		\mathcal{L}_{1;ij;i'j'}&=w_{ij}\delta_{ii'}\delta_{jj'}+\frac{J}{2D}\delta_{ij}\delta_{i'j'}+\frac{J}{2D}\delta_{ij'}\delta_{ji'},\nn
		w_{ij}&=-\ii E_{i}+\ii E_{j}-{1+D\over 2D}J\ed 
	\end{align}
	To solve the time evolution of the system, we make an ansatz 
	\begin{align}
		\mathcal{L}^n_{1;ij;i'j'}=c_{n}^{(ij)}\adjincludegraphics[valign=c, height=\diah\textwidth]{id2.pdf}+d_{n}^{(ii')}\adjincludegraphics[valign=c, height=\diah\textwidth]{cross2.pdf}+g_{n}^{(ij)}\adjincludegraphics[valign=c,height=\diah\textwidth]{exchange.pdf}\ed\nonumber
	\end{align}
	As we did for the GUE noise, we attempt to solve these unknown coefficients by utilizing their recursion relations. It is straightforward to find
	\begin{equation}
		\begin{aligned}
			\mathcal{L}^{n+1}_{1;ij;i'j'}&=w_{ij}c_{n}^{(ij)}\adjincludegraphics[valign=c, height=\diah\textwidth]{id2.pdf}
			+w_{ii}d_{n}^{(ii')}\adjincludegraphics[valign=c, height=\diah\textwidth]{cross2.pdf}\nn
			&\nl+w_{ij}g_{n}^{(ij)}\adjincludegraphics[valign=c, height=\diah\textwidth]{exchange.pdf} +{J\over 2D}c_n^{(i'i')}\adjincludegraphics[valign=c, height=\diah\textwidth]{cross2.pdf}\nn
			&\nl+{J\over 2D}\sum_k d_n^{(ki')}\adjincludegraphics[valign=c, height=\diah\textwidth]{cross2.pdf} 
			+{J\over 2D}g_n^{(i'i')}\adjincludegraphics[valign=c, height=\diah\textwidth]{cross2.pdf}\nn
			&\nl+{J\over 2D}c_n^{(ji)}\adjincludegraphics[valign=c, height=\diah\textwidth]{exchange.pdf} 
			+{J\over 2D} d_n^{(ii')}\adjincludegraphics[valign=c, height=\diah\textwidth]{cross2.pdf}\nn
			&\nl
			+{J\over 2D}g_n^{(ji)}\adjincludegraphics[valign=c, height=\diah\textwidth]{id2.pdf}
		\end{aligned}
	\end{equation}
	where we have neglected the index labels in the first expansion. 
	Comparing both sides of the equation and using the ansatz, we have 
	\begin{equation}
		\begin{aligned}
			c_{n+1}^{(ij)}&=w_{ij}c_{n}^{(ij)}+\frac{J}{2D}g_{n}^{(ji)},~g_{n+1}^{(ij)}=w_{ij}g_{n}^{(ij)}+\frac{J}{2D}c_{n}^{(ji)}\co \\
			d_{n+1}^{(ii')}&=w_{ii}d_{n}^{(ii')}+\frac{J}{2D}c_{n}^{(i'i')}+\frac{J}{2D}\sum_{k}d_{n}^{(ki')}\nn
			&\ns+\frac{J}{2D}g_{n}^{(i'i')}+\frac{J}{2D}d_{n}^{(ii')}\ed
		\end{aligned}
		\label{eq:GOEcoeff-recur}
	\end{equation}
	After solving the recursion relations, one can obtain the explicit expression of $\mathcal{U}_1$, we leave the detailed derivation in Appendix \ref{appdix:goe}. For the strong noise case $J\gg |E_{ij}|,\forall i,j$, we finally have 
		\begin{equation}
			\begin{aligned}
				\mathcal{U}_{1}(t)&=\frac{1}{2}\left(e^{-\frac{J}{2}t}+e^{-\frac{J}{2}t-\frac{J}{D}t}\right)\delta_{ii'}\delta_{jj'}
				+\frac{1}{D}\left(1-e^{-\frac{J}{2}t}\right)\delta_{ij}\delta_{i'j'}\nn
				&\ns+\frac{1}{2}\left(e^{-\frac{J}{2}t}-e^{-\frac{J}{2}t-\frac{J}{D}t}\right)\delta_{ij'}\delta_{ji'}\ed 
			\end{aligned}
			\label{eq:GOE-U1-Large-J}
		\end{equation}
	%
	Then the noise averaged SFF  is 
	\begin{align}
		K_{J\gg E}(t)&\sim\frac{1}{2}\left(e^{-\frac{J}{2}t}+e^{-\frac{J}{2}t-\frac{J}{D}t}\right)+\frac{1}{D^{2}}\left(1-e^{-\frac{J}{2}t}\right)\nn
		&\nl+\frac{1}{2D}\left(e^{-\frac{J}{2}t}-e^{-\frac{J}{2}t-\frac{J}{D}t}\right)\ed 
	\end{align}
	And the transition probability is 
	\begin{align}
		P_{i'\leftarrow i}(t)=\delta_{ii'}e^{-\frac{J}{2}t}+\frac{1}{D}\left(1-e^{-\frac{J}{2}t}\right)\ed 
	\end{align}
	Comparing with the GUE case, we find they are the same by a scaling $J\to 2 J$.

	For the general $J$ we finally have 
		\begin{equation}
			\begin{aligned}
				\mathcal{U}_{1}(t)&=\frac{1}{2}\left(c_{ij}^{-}e^{z_{ij}^{-}t}+c_{ij}^{+}e^{z_{ij}^{+}t}\right)\delta_{ii'}\delta_{jj'}+\frac{1}{D}\left(1-e^{-\frac{J}{2}t}\right)\delta_{ij}\delta_{i'j'}\nn
				&+\frac{1}{2}\left(g_{ij}e^{z_{ij}^{+}t}-g_{ij}e^{z_{ij}^{-}t}\right)\delta_{ij'}\delta_{ji'}
			\end{aligned}
			\label{eq:GOE-U1}
		\end{equation}
	%
	where we have defined 
	\begin{equation}
		\begin{aligned}
			g_{ij}&\equiv\frac{J}{D}\frac{1}{\sqrt{\left(w_{ij}-w_{ji}\right){}^{2}+J^{2}D^{-2}}}\co\nn
			c_{ij}^{\pm}&\equiv \frac{\sqrt{\left(w_{ij}-w_{ji}\right){}^{2}+J^{2}D^{-2}}\pm\left(w_{ij}-w_{ji}\right)}{\sqrt{\left(w_{ij}-w_{ji}\right){}^{2}+J^{2}D^{-2}}},\nn
			z_{ij}^{\pm}&\equiv \frac{1}{2}\left[\left(w_{ij}+w_{ji}\right)\pm\sqrt{\left(w_{ij}-w_{ji}\right){}^{2}+J^{2}D^{-2}}\right]\ed
		\end{aligned}
	\end{equation}
	\begin{figure}[ht]
		\begin{center}
			\includegraphics[width=0.43\textwidth]{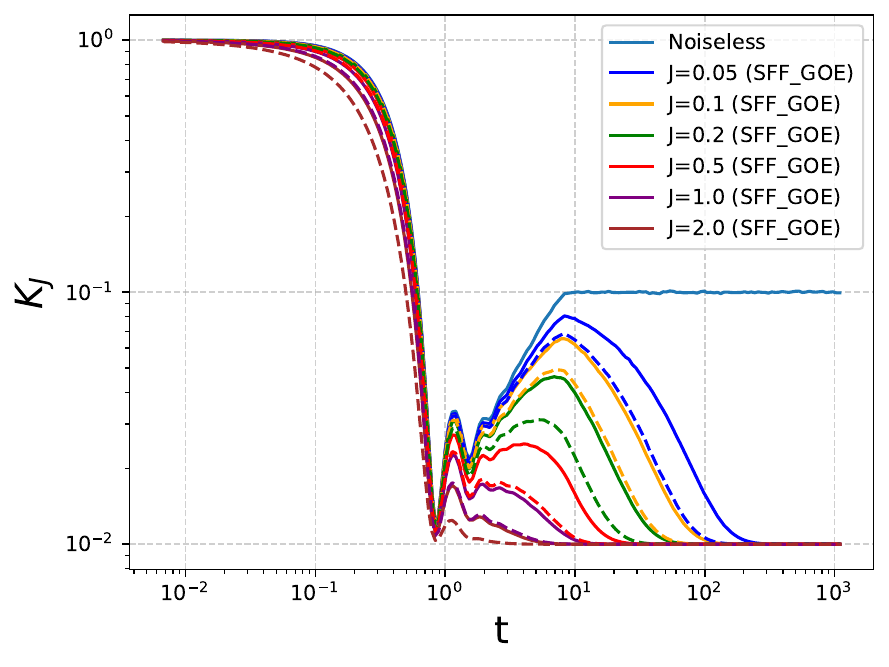}
			\caption{The effect of GOE noise on SFF, where the spectrum \(\{E_i\}_{i=1}^{D=10}\) is drawn from energy levels of  GUE. The plot is generated using Eq.~\eref{eq:GOE-sff} after performing an ensemble average over 40,000 realizations of \(\{E_i\}_{i=1}^{D=10}\). The dashed lines of the same color correspond to the GUE noise case for comparison.}
			\label{fig:GOE_SFFJ}
		\end{center}
	\end{figure}
	\paragraph{SFF}
	It is direct to calculate the noise-averaged SFF 
	\begin{align}\label{eq:GOE-sff}
		K_{J}(t)&=
		\frac{1}{D^{2}}\sum_{ij}\frac{1}{2}\left(c_{ij}^{-}e^{z_{ij}^{-}t}+c_{ij}^{+}e^{z_{ij}^{+}t}\right)+\frac{1}{D^{2}}\left(1-e^{-\frac{J}{2}t}\right)\nn
		&\nl+{1\over D}e^{-\frac{(D+1)J}{2D}t}\sinh\left(\frac{J}{2D}t\right)\ed
	\end{align}
	Here, we observe that the noise-averaged SFF consists of two components: one that depends on the spectrum and another that is universal. And when $t\to \infty$, the noise-averaged SFF approach ${1/D^2}$ which is the same as GUE noise.  Unlike the GUE noise, for general $J$, there is no clear relationship between the noise-averaged SFF \( K_J(t) \) and the noiseless SFF \( K_{J=0}(t) \). For small \( J\ll |E_{ij}|,\forall i,j \), we have 
	\begin{align}
		K_{J\ll E}(t)&\sim e^{-\frac{D+1}{2D}Jt}K_{J=0}(t)-\frac{1}{D}e^{-\frac{D+1}{2D}Jt}\cosh\left(\frac{Jt}{2D}\right)\nn
		&\nl+\frac{1}{D^{2}}\left(1-e^{-\frac{J}{2}t}\right)+{1\over D}e^{-\frac{D+1}{2D}Jt}\sinh\left(\frac{Jt}{2D}\right)\ed
	\end{align}
    For large $J$ expansion, we have 
    \begin{align}
    	K_{J\gg E}(t)&\sim e^{-\frac{D+1}{2D}Jt}\cosh\left(\frac{Jt}{2D}\right)+\frac{1}{D^{2}}\left(1-e^{-\frac{J}{2}t}\right)\nn
    	&\nl +\frac{1}{D}e^{-\frac{D+1}{2D}Jt}\sinh\left(\frac{Jt}{2D}\right)\ed
    \end{align}
     Comparing with the GUE noise Eq.\eref{eq:GUE-sff}, for large $D$,  we find the suppressed factor is 
    \begin{align}
    	f_{\text{GOE}}=e^{-Jt/2},~f_{\text{GUE}}=e^{-Jt}
    \end{align}
which means the GOE noise-average SFF decays slower than the GUE case, as shown in Fig.~\ref{fig:GOE_SFFJ}.
    \paragraph{Two-point function}
	The noise-averaged two-point function for GOE noise is given by
	\begin{align}
		C_J(t)=&\frac{1}{2D}\sum_{ij}\left(c_{ij}^{-}e^{z_{ij}^{-}t}+c_{ij}^{+}e^{z_{ij}^{+}t}\right)O^\dagger_{ij}O_{ji}\nn
		&+\frac{1}{D^2}\left(1-e^{-\frac{J}{2}t}\right)\text{Tr}O^\dagger\text{Tr}O\nn
		&+\frac{1}{2D}\sum_{ij}\left(g_{ij}e^{z_{ij}^{+}t}-g_{ij}e^{z_{ij}^{-}t}\right)O^\dagger_{ji}O_{ji}\ed
	\end{align}
    The formula appears more complicated than the GUE noise case, but it remains convenient for numerical calculations. For small $J$, we have 
    \begin{align}
         C_{J\ll E}(t)=&e^{-\frac{D+1}{2D}Jt}C_{J=0}(t)+\left(1-e^{-\frac{J}{2}t}\right){\text{Tr}O^{\dagger}\text{Tr}O\over D^2}\nn
         &-\frac{Je^{-\frac{D+1}{2D}Jt}}{2D^2}\sum_{i\not=j}\frac{\sin\left(|E_{ij}|t\right)}{|E_{ij}|}O_{ji}^{\dagger}O_{ji}\ed
    \end{align}
   Given that $J \ll 1$, the term in the second line can be neglected. By comparing with Eq.\eref{eq:2pt_GUE}, we observe that for large $D$, the noise-averaged $C_J(t)$
    in the GOE case can be derived from the GUE noise scenario by substituting $J$ with $\frac{1}{2}J$.
    
    For large $J$, we have
    \begin{align}
    	C_{J\gg E}(t)=&e^{-\frac{D+1}{2D}Jt}\cosh\left(\frac{Jt}{2D}\right){\text{Tr}\left(O^{\dagger}O\right)\over D}\nn
    	&+\left(1-e^{-\frac{J}{2}t}\right){\text{Tr}O^{\dagger}\text{Tr}O\over D^2}\nn
    	&+e^{-\frac{D+1}{2D}Jt}\sinh\left(\frac{Jt}{2D}\right){\text{Tr}\left(O^{2}\right)\over D}\ed
    \end{align}
    Additionally, one can assess the effects of GOE noise on Krylov complexity. We will not conduct that calculation here.

	\subsection{General case}
	\label{sec:GOE-general}
	Similar to our approach for GUE noise, we can propose an ansatz of the form
	\begin{equation}
		\begin{aligned}
			\mathcal{U}_{1;ij;i'j'}(t)=C_{ij}(t)\adjincludegraphics[valign=c, height=\diah\textwidth]{id2.pdf}+D_{ii'}(t)\adjincludegraphics[valign=c, height=\diah\textwidth]{cross2.pdf}+
			G_{ii'}(t) \adjincludegraphics[valign=c, height=\diah\textwidth]{exchange.pdf}
		\end{aligned}
		\label{eq:U1}
	\end{equation}
	where $C_{ij},D_{ii'},G_{ii'}$ are three functions to be determined. Using the equation of motion for $\mathcal{U}_2$, 
	we have 
	\begin{equation}
		\begin{aligned}
			\ddt C_{ij}&=w_{ij}C_{ij}+\frac{1}{2}\lambda_{ij}G_{ji},\ddt G_{ii'}=w_{ii'}G_{ii'}+\frac{1}{2}\lambda_{ii'}C_{i'i}\\
			\ddt D_{ii'}&=w_{ii}D_{ii'}+\frac{1}{2}\lambda_{ii'}C_{i'i'}+\frac{1}{2}\sum_{s}\lambda_{is}D_{si'}\\
			&\nl+\frac{1}{2}\lambda_{ii'}G_{i'i'}+\frac{1}{2}\lambda_{ii}D_{ii'}\ed
		\end{aligned}
	\end{equation}
	Here 
	\begin{align}
		w_{ij}=-\ii E_i + \ii E_j-\frac{1}{4}\left(J_{i}+J_{j}+\lambda_{ii}+\lambda_{jj}\right)\ed
	\end{align}One can first solve $C,G$  by setting 
	\begin{align}
		C_{ij}(t)=c_{ij}(t)\exp\left(w_{ij}t\right),G_{ii'}(t)=g_{ii'}(t)\exp\left(w_{ii'}t\right)\co\nonumber
	\end{align}
	then we have 
	\begin{align}
		\ddtt c_{ij}+\left(w_{ij}-w_{ji}\right)\ddt c_{ij}=\frac{1}{4}\lambda_{ij}\lambda_{ji}c_{ij}
	\end{align}
	with the initial condition $c_{ij}(0)=1,c_{ij}'(0)=0$, we finally have 
	\begin{align}
		C_{ij}=\frac{1}{2}\left(c_{ij}^{-}e^{z_{ij}^{-}t}+c_{ij}^{+}e^{z_{ij}^{+}t}\right),G_{ij}=\frac{1}{2}\left(g_{ij}e^{z_{ij}^{+}t}-g_{ij}e^{z_{ij}^{-}t}\right)
	\end{align}
	where we have defined 
	\begin{equation}
		\begin{aligned}
			g_{ij}&=\frac{\lambda_{ij}}{\sqrt{\left(w_{ij}-w_{ji}\right){}^{2}+\lambda_{ij}^2}}\co\\
			c_{ij}^{\pm}&=\frac{\sqrt{\left(w_{ij}-w_{ji}\right){}^{2}+\lambda_{ij}^2}\pm\left(w_{ij}-w_{ji}\right)}{\sqrt{\left(w_{ij}-w_{ji}\right){}^{2}+\lambda_{ij}^2}}\co\\
			z_{ij}^{\pm}&=\frac{1}{2}\left[\left(w_{ij}+w_{ji}\right)\pm\sqrt{\left(w_{ij}-w_{ji}\right){}^{2}+\lambda_{ij}^2}\right]\ed
		\end{aligned}
	\end{equation}
    If we take \(\lambda_{ij} = \frac{J}{D}\), we see that the definition returns to the constant case. We then deal with \(D_{ii'}\), noting that only the diagonal parts of \(C\) and \(G\) appear in the equation for \(D\). If we take \(i = j\), we have \(g_{ii} = 1\), \(c_{ii}^{\pm} = 1\), and \(z_{ii}^{\pm} = {-J_i-\lambda_{ii}\over 2} \pm {\lambda_{ii}\over 2}\). Define

	\begin{align}
		P_{ij}=-{J_i\over 2}\delta_{ij}+{\lambda_{ij}\over 2},Q_{ii'}(t)={\lambda_{ii'}\over 2}e^{{-J_i\over 2}t}\co
	\end{align}
    then we have 
	\begin{align}
		D_{ii'}(t)=\left[\exp\left(Pt\right).\int_{0}^{t}dt\exp\left(P^{-1}t\right).Q(t)\right]_{ii'}\ed
	\end{align}
    This gives us an explicit expression for $\mathcal{U}_1$. We can then study the effects of noise as before.
	\section{Two replica observables: GUE noise}
	\label{sec:U2}
	To study the effect of noise on another quantum chaos diagnostic, OTOC, we need to consider the time evolution of two operators, which means we need to compute $\mathcal{U}_2$. As discussed in the paper, we can obtain the expression of $\mathcal{L}_2$ by taking noise average and keep the linear contribution

	\begin{equation}
		\begin{aligned}
			&\mathbb{I}^{\otimes 4}+\mathcal{L}_{2}\Delta t\\
			=&\mathbb{E}\Bigg\{\left[\mathbb{I}-\ii H\Delta t-\frac{1}{2}H^{2}\Delta t^{2}\right]\otimes\left [\mathbb{I}+\ii H^{*}\Delta t-\frac{1}{2}H^{*2}\Delta t^{2}\right]\\
			&\otimes\left [\mathbb{I}-\ii H\Delta t-\frac{1}{2}H^{2}\Delta t\right] \otimes\left [\mathbb{I}+\ii H^{*}\Delta t-\frac{1}{2}H^{*2}\Delta t^2\right]\Bigg\}\co
		\end{aligned}
	\end{equation}
	then we obtain
	\begin{align}
		\mathcal{L}_2&=w_{ijkl}\delta_{ijkl;i'j'k'l'}+\frac{J}{D}\left(P_{1\bar{1}}+P_{2\bar{2}}+P_{1\bar{2}}+P_{2\bar{1}}\right)\nn
		&\nl-\frac{J}{D}\left(X_{12}+X_{\bar{1}\bar{2}}\right)
	\end{align}
	where \( I \) denotes the identity operator, and the other two operators are represented below
	\begin{align}
		P_{i\bar{j}}=\adjincludegraphics[valign=c, height=\diah\textwidth]{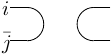},X_{ij}=\adjincludegraphics[valign=c,height=\diah\textwidth]{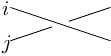}\ed
	\end{align}   
    Here we use $i,j$ to  label the channel of $U_t$ while $\bar{i},\bar{j}$ to label the channels of $U_t^*$. 
	We use the graph representation
	\begin{align}
		\mathcal{L}_2= w_{ijkl}\mathsf{F}_{11}+{J\over D}(\mathsf{F}_{21}+\mathsf{F}_{22}+\mathsf{F}_{23}+\mathsf{F}_{24})-{J\over D}(\mathsf{F}_{31}+\mathsf{F}_{32})
	\end{align}
	where we have ignored the index labels \( ijkl; i'j'k'l' \) and we have defined.
	\begin{align}
		w_{ijkl}=-\ii E_i+\ii E_j-\ii E_k+\ii E_l-2J\ed
	\end{align}
	If \( E_i = 0 \), then everything becomes the same as \cite{Tang:2024kpv}.
	\begin{figure}[h]
		\begin{center}
			\includegraphics[width=0.45\textwidth]{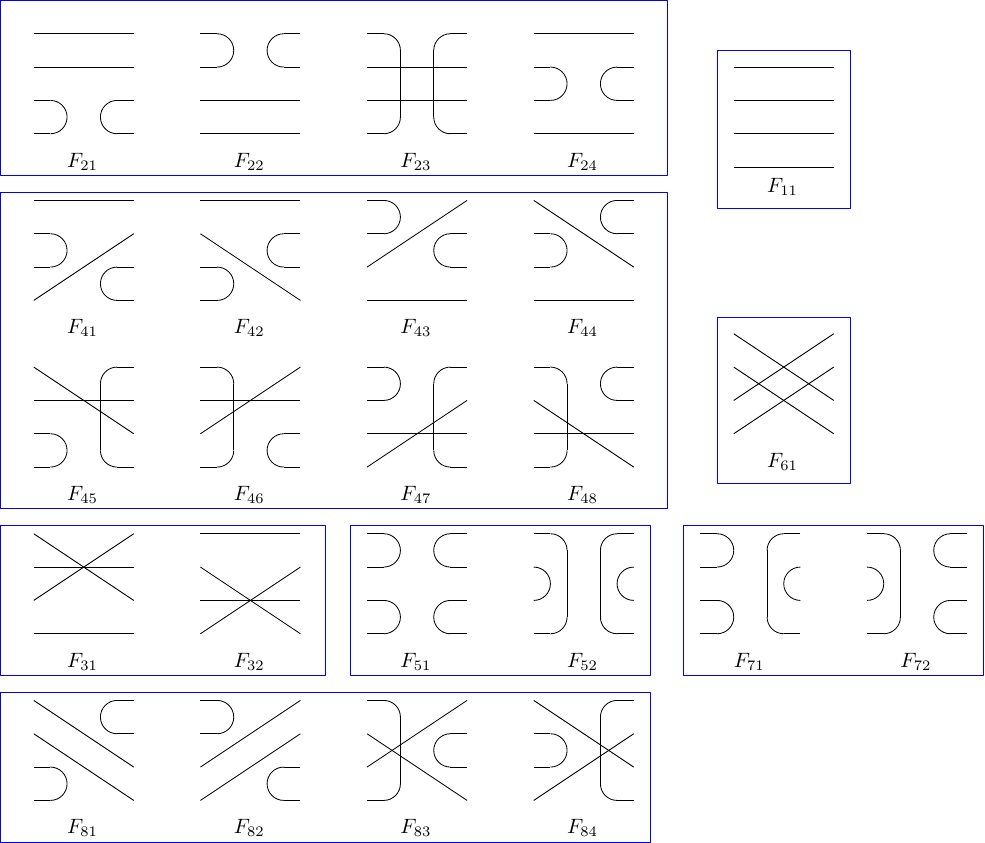}
			\caption{8 groups of graphs for $k=2$.}
			\label{fig:k2Fs}
		\end{center}
	\end{figure}
	As depicted in Fig.\ref{fig:k2Fs}, there are \( 24 \) kinds of graphs for \( k = 2 \), so in principle, we will need to deal with \( 24 \) unknown coefficients. It could be a tough task to find their recursion relations and solve them.
	
	Here, we need to notice the symmetry of \( w_{ijkl} \); it is invariant under the exchange of \( i, k \) and \( j, l \). Identifying any pair of \( i, \bar{j} \) (setting \( E_i - E_{\bar{j}} = 0 \)) leads to no dependence on them. It is easy to see that the factor \( w_{ijkl} \) commutes with the graphs in \( \mathcal{L}_2 \), since all graphs are generated by \( \mathcal{L}_2 \), so \( w_{ijkl} \) commutes with all graphs.
	\begin{align}
		w_{ijkl}\mathsf{F}_{\alpha}= \mathsf{F}_{\alpha} w_{i'j'k'l'}\ed
	\end{align}
    Consequently, we can consistently position \( w_{ijkl} \) to the right of the graphs, denoting it simply as a constant \( w \). Therefore, our focus shifts to calculating the combination law of the graphs
	\begin{align}
		\mathsf{F}_{\alpha}.\mathsf{F}_{\beta}=\mathsf{F}_{(\alpha,\beta)}\ed
	\end{align}
	Here, \( \mathsf{F}_{\alpha} \) belongs to \( \mathcal{L}_2 \), while \( \mathsf{F}_{\beta} \) can be any graph generated by \( \mathcal{L}_2 \). Therefore, we need to study the action of \( \mathcal{L}_2 \) on the 24 individual graphs, \textit{i}.\textit{e}., \( \mathcal{L}_2 . \mathsf{F}_{\alpha} \). It can be observed that these 24 graphs can be divided into 8 groups \( \mathsf{F}_a \equiv \sum_b \mathsf{F}_{ab} \), as depicted in Fig.\eref{fig:k2Fs}. 
	Furthermore, we have \( \mathcal{L}_2 . \mathsf{F}_a = M_{ba} \mathsf{F}_b \), where \( M \) is an \( 8 \times 8 \) matrix:
		\begin{align}
			\left(
			\begin{array}{cccccccc}
				w & 0 & -\frac{2 J}{D} & 0 & 0 & 0 & 0 & 0 \\
				\frac{J}{D} & J+w & 0 & 0 & 0 & 0 & 0 & 0 \\
				-\frac{J}{D} & 0 & w & 0 & 0 & -\frac{J}{D} & 0 & 0 \\
				0 & 0 & \frac{J}{D} & J+w & 0 & 0 & 0 & 0 \\
				0 & \frac{2 J}{D} & 0 & 0 & 2 J+w & 0 & 0 & \frac{2J}{D} \\
				0 & 0 & -\frac{2 J}{D} & 0 & 0 & w & 0 & 0 \\
				0 & 0 & 0 & \frac{4 J}{D} & 0 & 0 & 2 J+w & 0 \\
				0 & 0 & 0 & 0 & 0 & \frac{J}{D} & 0 & J+w \\
			\end{array}
			\right)\ed
		\end{align}
	We can use the same method as in \cite{Tang:2024kpv} to solve $\mathcal{U}_2$
	\begin{align}
		\mathcal{U}_2(t)=e^{-\ii E_{ijkl }t}\sum_{i=1}^{8}f_a(t)\mathsf{F}_a
	\end{align}
	where $E_{ijkl}\equiv E_i-E_j+E_k-E_l$ and 
	\begin{widetext}
	\begin{align}
		\begin{bmatrix}f_{1}(t)\\
			f_{2}(t)\\
			f_{3}(t)\\
			f_{4}(t)\\
			f_{5}(t)\\
			f_{6}(t)\\
			f_{7}(t)\\
			f_{8}(t)
		\end{bmatrix}=\begin{bmatrix}0 & 0 & \frac{1}{4} & \frac{1}{2} & \frac{1}{4}\\
			0 & \frac{D^{2}-2}{D\left(D^{2}-4\right)} & \frac{-1}{4(D-2)} & \frac{-1}{2D} & \frac{-1}{4(D+2)}\\
			0 & 0 & -\frac{1}{4} & 0 & \frac{1}{4}\\
			0 & \frac{-1}{D^{2}-4} & \frac{1}{4(D-2)} & 0 & \frac{-1}{4(D+2)}\\
			\frac{1}{D^{2}-1} & \frac{-2}{D^{2}-4} & \frac{1}{2(D-1)(D-2)} & 0 & \frac{1}{2(D+1)(D+2)}\\
			0 & 0 & \frac{1}{4} & -\frac{1}{2} & \frac{1}{4}\\
			\frac{-1}{D^{3}-D} & \frac{4}{D(D^{2}-4)} & \frac{-1}{2(D-1)(D-2)} & 0 & \frac{1}{2(D+1)(D+2)}\\
			0 & \frac{2}{D(D^2-4)} & \frac{-1}{4(D-2)} & \frac{1}{2D} & \frac{-1}{4(D+2)}
		\end{bmatrix}\times\begin{bmatrix}1\\
			e^{-Jt}\\
			e^{-(2-2D^{-1})Jt}\\
			e^{-2Jt}\\
			e^{-(2+2D^{-1})Jt}
		\end{bmatrix}\ed
	\end{align}
\end{widetext}
	With the expression of $\mathcal{U}_2$, we can calculate the fluctuation of the two-point function and SFF. For example, the variance of SFF is given by
	\begin{align}
		\mathbb{E}\left(\text{Tr}U_{t}\text{Tr}U_{t}^{\dagger}\text{Tr}U_{t}\text{Tr}U_{t}^{\dagger}\right)=\sum_{a=1}^8f_{a}V_a(\text{SFF})
	\end{align}
	where $\{V_a(\text{SFF})\}_{a=1}^8$ is a vector 
	\begin{widetext}
	\begin{equation}
		\begin{aligned}
			&\left(D^{4}K(t)^{2},4D^{3}K(t),
		\text{Tr}U_{2t}^{(0)}\left(\text{Tr}U_{t}^{(0)\dagger}\right)^{2}+\text{Tr}U_{2t}^{(0)\dagger}\left(\text{Tr}U_{t}^{(0)}\right)^{2},
			8D^{2}K(t),2D^{2},D^{2}K(2t),2D,4D\right)^T\ed
		\end{aligned}
	\end{equation}
	\end{widetext}
	Moreover, we can investigate the noise effect on the OTOC at infinite temperature. Consider two operators \( A \) and \( B \), and define \( B_t \equiv U_t^\dagger B U_t \). We have
	\begin{equation}
		\begin{aligned}
			\text{OTOC}_{J}&=\sum_{a=1}^8 f_a(t)
			\adjincludegraphics[valign=c, height=0.05\textwidth]{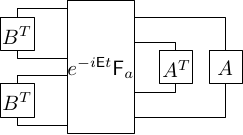}
		\end{aligned}
	\end{equation}
	where we have defined $\text{OTOC}_{J}\equiv \frac{1}{D}\text{Tr}\mathbb{E}\left(AB_{t}AB_{t}\right)$, $e^{-\ii \mathsf{E}t}\mathsf{F}_a=\sum_be^{-\ii \mathsf{E}t}\mathsf{F}_{ab}$  are 8 graph categories times the spectral information factor $e^{-\ii \mathsf{E}t}=e^{-\ii E_{ijkl}t}$ from left-hand side  or $e^{-\ii \mathsf{E}t}=e^{-\ii E_{i'j'k'l'}t}$ from the right-hand side. One can check the disorder-free OTOC  can be obtained by taking $J=0$, \textit{i}.\textit{e}., $\frac{1}{D}\text{Tr}\left(AB_{t}^{(0)}AB_{t}^{(0)}\right)=\text{OTOC}_{J=0}$. 
	For simplicity, we consider the case where \( \text{Tr} A = \text{Tr} B = 0 \), which implies that only \( \mathsf{F}_1, \mathsf{F}_3, \mathsf{F}_6 \) contribute. Thus, we have

	\begin{widetext}
		\begin{align}
			\text{OTOC}_{J}&=\left(f_{1}(t)+f_{6}(t)\right)\text{OTOC}_{J=0}+\frac{2}{D}f_{3}(t)\text{Tr}\left(AB_{t}^{(0)}\right)\text{Tr}\left(AB^{(0)}_{t}\right)\nn
			&=e^{-2Jt}\left[\cosh\left(2D^{-1}t\right)\text{OTOC}_{J=0}+\sinh\left(2D^{-1}t\right)\frac{1}{D}\text{Tr}\left(AB_{t}^{(0)}\right)\text{Tr}\left(AB^{(0)}_{t}\right)\right]\ed
		\end{align}
	\end{widetext}
	For large \( D \), we can neglect the terms suppressed by \( \frac{1}{D} \) and obtain
	\begin{align}\label{eq:otoc_GUE}
		\text{OTOC}_{J}=e^{-2Jt}\text{OTOC}_{J=0}+O(1/D)\ed
	\end{align}
	While quantum chaos is characterized by exponential decay at early times, noise enhances this decay by an exponential factor $e^{-2Jt}$. Consequently, the diagnostic capability of the OTOC diminishes when \( J \) is relatively large.
	\section{Conclusion and Outlook}\label{sec:conclusion}
	This study investigates the impact of noise on the diagnostics of quantum chaos, specifically focusing on the spectral form factor (SFF), Krylov complexity, and out-of-time correlators (OTOCs). The findings reveal that the presence of noise, particularly in the form of white noise, introduces significant challenges in accurately diagnosing quantum chaos. The results indicate that as the strength of noise increases, the effectiveness of these quantum chaos diagnostics diminishes. 
	
	However, it seems the $r$-parameter still serves as a good diagnostic for quantum chaos in the case of GUE noise with constant $\lambda_{ij}$, provided the effective Hamiltonian defined in Eq.~\eqref{eq:Heff} can be measured experimentally. This remarkable result stems from the fact that the new energy levels $E_{J;i}$ are linear functions of $E_i$ (Eq.~\eqref{eq:newEi}), ensuring that the $r_n$ remains invariant. As discussed in the main text, since any measurement requires a finite time $\Delta t > 0$ and the level spacings $s_n$ are exponentially suppressed, the $r$-parameter will lose its validity for large values of $J\Delta t$.
	
	For other cases considered in this paper, we can obtain a similar relationship by evaluating the contraction with $\mathcal{U}_1$ (see Eq.~\eqref{eq:Heff-fig})
	\begin{align}
		E_{J;i} = \alpha_i E_{i} + c_i\ed
	\end{align}
	For GUE/GOE noise with constant $\lambda_{ij}$, the coefficients $\alpha_i$ and $c_i$ are universal (independent of the index $i$). However, for general $\lambda_{ij}$, these coefficients depend on both $\lambda_{ij}$ and $E_i$, meaning $\alpha_i$ and $c_i$ vary accordingly. Since the expression for $\mathcal{U}_1$ involves matrix inversions and integrations, it is not straightforward to evaluate them explicitly. For small values of $\frac{1}{D}\sum_{ij}\lambda_{ij}\Delta t$, the time evolution is expected to be nearly unitary, and the $r$-parameter remains effective.

	It is also interesting to consider other kinds of noise like GSE noise and general replica dynamics. 
	But things can be more complex. Taking replica-2 dynamics with GOE noise as  an example, there are more graphs in the operator $\mathcal{L}_2$. Moreover, the weight factor  $w_{ijkl}$ does not commutes with all graphs in $\mathcal{L}_2$, which leads to the calculation extremely complex.  
	
	The implications of this study are profound for future research in quantum chaos and open quantum systems. The results underscore the necessity for developing more robust diagnostic tools that can account for environmental influences such as noise. As real-world systems are invariably open and subject to various forms of perturbations, understanding how these factors interact with quantum chaos diagnostics is crucial. Future investigations could explore alternative forms of noise or different types of quantum systems to further elucidate these dynamics. Additionally, examining how these findings relate to practical applications in quantum computing and information processing could yield significant insights into enhancing system resilience against environmental disturbances. In conclusion, while traditional diagnostics for quantum chaos have provided valuable insights into chaotic behavior, their efficacy is notably diminished in noisy environments. This study highlights the critical need for ongoing research to adapt and refine these tools in light of environmental complexities inherent in real-world quantum systems.
	
	\begin{acknowledgments}
	I would like to thank Professor Cheng Peng for his support regarding my accommodation, as well as for the discussions with Yingyu Yang, Yanyuan Li and Miao Wang.  TL is supported by NSFC NO. 12175237.
	\end{acknowledgments}
	\appendix 
	\section{Solving the recursion relations}
	\label{appdix:goe}
    As discussed in the main text, we encounter a recursion relation when dealing with \( \mathcal{U}_1 \) for a system with GOE noise
	\begin{align}\label{eq:GOEcoeff-recur-copy}
		c_{n+1}^{(ij)}&=w_{ij}c_{n}^{(ij)}+\frac{J}{2D}g_{n}^{(ji)}\co~ g_{n+1}^{(ij)}=w_{ij}g_{n}^{(ij)}+\frac{J}{2D}c_{n}^{(ji)}\co \nn
		d_{n+1}^{(ii')}&=w_{ii}d_{n}^{(ii')}+\frac{J}{2D}c_{n}^{(i'i')}+\frac{J}{2D}\sum_{k}d_{n}^{(ki')}\nn
		&\nl+\frac{J}{2D}g_{n}^{(i'i')}+\frac{J}{2D}d_{n}^{(ii')}\ed
	\end{align}
	One can observe that the recursion relations for \( c_n^{(ij)} \) and \( g_n^{(ij)} \) do not involve \( d_n^{(ii')} \), allowing us to tackle the solutions for \( c_n^{(ij)} \) and \( g_n^{(ij)} \) first. Below, we list some leading terms
	\begin{equation}
		\begin{aligned}
			c_1^{(ij)}& = w_{ij}, g_1^{(ij)}= {J\over 2D}\co \\
			c_2^{(ij)}& = w_{ij}^2+\frac{J^2}{4 D^2}, g_2^{(ij)}=\frac{J \left(w_{ij}+w_{ji}\right)}{2 D}\co  \\
			c_3^{(ij)}& = \frac{4 D^2 w_{ij}^3+2 J^2 w_{ij}+J^2 w_{ji}}{4 D^2},\\
			 g_3^{(ij)}&= \frac{J\left(4D^{2}w_{ij}^{2}+4D^{2}w_{ji}^{2}+4D^{2}w_{ij}w_{ji}+J^{2}\right)}{8D^{3}}\ed 
		\end{aligned}
	\end{equation}
	\subsection{Strong noise limit}
	For simplicity, we first consider the strong noise case \( J \gg |E_{ij}|,\forall i,j \), leading to \( w_{ij} = -\frac{D+1}{2D} J \). In this scenario, we find that all structure coefficients \( c_n^{(ij)} \) and \( g_n^{(ij)} \) do not depend on \( i \) or \( j \). Thus, we can drop the indices, setting \( c_n = c_n^{(ij)} \). We then find
	\begin{align}
		c_{n}&=\frac{1}{2}\left[\left(-\frac{J}{2}\right)^{n}+\left(-\frac{J}{2}-\frac{J}{D}\right)^{n}\right]\co \\
		g_{n}&=\frac{1}{2}\left[\left(-\frac{J}{2}\right)^{n}-\left(-\frac{J}{2}-\frac{J}{D}\right)^{n}\right]\ed 
	\end{align}
	Plugging the result into the equation for \( d_n \), we find
	\begin{equation}
		\begin{aligned}
			d_{n+1}&=\frac{J}{2D}\left(\frac{-J}{2}\right)^{n}\ed
		\end{aligned}
	\end{equation}
	So we can construct $\mathcal{U}_1$ directly
	\begin{align}
		\mathcal{U}_{1}(t)&=\delta_{ii'}\delta_{jj'}+\sum_{n=1}^\infty\frac{\mathcal{L}_{1;ij;i'j'}^{n}t^{n}}{n!}\nn
		&=\delta_{ii'}\delta_{jj'}+\sum_{n=1}^\infty\frac{c_{n}t^{n}}{n!}\delta_{ii'}\delta_{jj'}\nn
		&\nl+\sum_{n=1}\frac{d_{n}t^{n}}{n!}\delta_{ij}\delta_{i'j'}+\sum_{n=1}^\infty\frac{g_{n}t^{n}}{n!}\delta_{ij'}\delta_{ji'}\nn
		&=\frac{1}{2}\left(e^{-\frac{J}{2}t}+e^{-\frac{J}{2}t-\frac{J}{D}t}\right)\delta_{ii'}\delta_{jj'}+\frac{1}{D}\left(1-e^{-\frac{J}{2}t}\right)\delta_{ij}\delta_{i'j'}\nn
		&\nl+\frac{1}{2}\left(e^{-\frac{J}{2}t}-e^{-\frac{J}{2}t-\frac{J}{D}t}\right)\delta_{ij'}\delta_{ji'}\ed 
	\end{align}
	\subsection{General $J$}
	For the general $J$, we can introduce the generating functions
	\begin{align}
		C^{(ij)}:=\sum_{n=1}^{\infty}z^{n}c_{n}^{(ij)},G^{(ij)}:=\sum_{n=1}^{\infty}z^{n}g_{n}^{(ij)}\ed
	\end{align}
	Using Eq.\eref{eq:GOEcoeff-recur}, we find 
	\begin{align}
		C^{(ij)}-zc_{1}^{(ij)}&=z\left(w_{ij}C^{(ij)}+\frac{J}{2D}G^{(ji)}\right)\co \\ 
		G^{(ij)}-zg_{1}^{(ij)}&=z\left(w_{ij}G_{n}^{(ij)}+\frac{J}{2D}C^{(ji)}\right)\ed
	\end{align}
	We can first write $C^{(ij)}$ in terms of $G^{(ij)}$ 
	\begin{align}
		C^{(ij)}=\frac{zc_{1}^{(ij)}+\frac{Jz}{2D}G^{(ji)}}{1-zw_{ij}}\co
	\end{align}
	then 
	\begin{align}
		G^{(ij)}-zg_{1}^{(ij)}=z\left(w_{ij}G_{n}^{(ij)}+\frac{J}{2D}\frac{zc_{1}^{(ji)}+\frac{Jz}{2D}G^{(ij)}}{1-zw_{ji}}\right)\ed
	\end{align}
	Then we find
	\begin{equation}
		\begin{aligned}
			G^{(ij)}&=\frac{zg_{1}^{(ij)}+\frac{J}{2D}\frac{z^{2}c_{1}^{(ji)}}{1-zw_{ji}}}{1-zw_{ij}-\left(\frac{J}{2D}\right)^{2}\frac{z^{2}}{1-zw_{ji}}}\\
			&=\frac{\left(1-zw_{ji}\right)zg_{1}^{(ij)}+\frac{J}{2D}z^{2}c_{1}^{(ji)}}{\left(1-zw_{ij}\right)\left(1-zw_{ji}\right)-\left(\frac{J}{2D}\right)^{2}z^{2}}\\&=\frac{J}{2D}\frac{z}{\left(1-zw_{ij}\right)\left(1-zw_{ji}\right)-\left(\frac{J}{2D}\right)^{2}z^{2}}\ed
		\end{aligned}
	\end{equation}
	It is evident that \( G^{(ij)} \) is symmetric with respect to \( i \) and \( j \), as indicated in the expressions for the leading terms of \( g_n^{(ij)} \). In principle, we can determine the expressions for the expansion coefficients by taking residues
	\begin{align}
		g_{n}^{(ij)}&=\frac{1}{2\pi i}\oint_{z=0}\frac{1}{z^{n+1}}G^{(ij)}=\frac{-1}{2\pi i}\sum_{z_i\not=0}\oint_{z=z_{i}}\frac{1}{z^{n+1}}G^{(ij)}
	\end{align}
    where we change the integral contour so that we need to sum all residues at \( z = z_i \neq 0 \). Since we always have \( n \ge 1 \), there are no singularities at infinity. Thus, we only need to consider the singularities arising from the denominators of \( G^{(ij)} \) and \( C^{(ij)} \). Direct calculation yields
	\begin{align}
		g_{n}^{(ij)}&={1\over 2}g_{ij}\left(z_{ij}^{+}\right)^{n}-{1\over 2}g_{ij}\left(z_{ij}^{-}\right)^{n}\co\nn
		c_n^{(ij)}&={1\over 2}c_{ij}^+\left(z^+_{ij}\right)^{n}+{1\over 2}c_{ij}^-\left(z^-_{ij}\right)^{n}\co
	\end{align}
	where we have defined
	\begin{align}
		g_{ij}&=\frac{J}{D}\frac{1}{\sqrt{\left(w_{ij}-w_{ji}\right){}^{2}+J^{2}D^{-2}}}\co\nn
		c_{ij}^{\pm}&=\frac{\sqrt{\left(w_{ij}-w_{ji}\right){}^{2}+J^{2}D^{-2}}\pm\left(w_{ij}-w_{ji}\right)}{\sqrt{\left(w_{ij}-w_{ji}\right){}^{2}+J^{2}D^{-2}}}\co\nn
		z_{ij}^{\pm}&=\frac{1}{2}\left[\left(w_{ij}+w_{ji}\right)\pm\sqrt{\left(w_{ij}-w_{ji}\right){}^{2}+J^{2}D^{-2}}\right]\ed
	\end{align}
	When \( i = j \), \( c_{n}^{(ii)} \) and \( g_{n}^{(ii)} \) are the same as in the strong noise case. Therefore, the expression for \( d_n^{(ij)} \) is also the same as in the strong noise limit. Using these exact expressions, we can calculate \( \mathcal{U}_1 \)
	\begin{widetext}
		\begin{equation}
			\begin{aligned}
				\mathcal{U}_{1}(t)&=\delta_{ii'}\delta_{jj'}+\sum_{n=1}^{\infty}\frac{\mathcal{L}_{1;ij;i'j'}^{n}t^{n}}{n!}\\&=\delta_{ii'}\delta_{jj'}+\sum_{n=1}^\infty\frac{c_{n}^{(ij)}t^{n}}{n!}\delta_{ii'}\delta_{jj'}+\sum_{n=1}^\infty\frac{d_{n}^{(ii')}t^{n}}{n!}\delta_{ij}\delta_{i'j'}+\sum_{n=1}^\infty\frac{g_{n}^{(ij)}t^{n}}{n!}\delta_{ij'}\delta_{ji'}\\
				&=\frac{1}{2}\left(c_{ij}^{-}e^{z_{ij}^{-}t}+c_{ij}^{+}e^{z_{ij}^{+}t}\right)\delta_{ii'}\delta_{jj'}+\frac{1}{D}\left(1-e^{-\frac{J}{2}t}\right)\delta_{ij}\delta_{i'j'}+\frac{1}{2}\left(g_{ij}e^{z_{ij}^{+}t}-g_{ij}e^{z_{ij}^{-}t}\right)\delta_{ij'}\delta_{ji'}\ed
			\end{aligned}
			\label{eq:GOE-U1}
		\end{equation}
	\end{widetext}
	The expression of $g_{ij},c_{ij}^{\pm},z_{ij}^{\pm}$ seems complicated. For $i=j$, they simplifies to 
	\begin{align}
		z_{ii}^{\pm}=-\frac{(D+1)J}{2D}\pm\frac{J}{2D},~c_{ii}^{\pm}=g_{ii}=1\ed
	\end{align}
	For $i\not=j$, we consider small $J$ and large $J$ expansion.
	\paragraph{Small $J$ expansion}
	\begin{align}\label{eq:smallJexp}
		g_{ij}&=\frac{-\ii J}{2 D|E_{ij}|}+O\left(J^{3}\right)\co\nn
		c_{ij}^{\pm}&=\left(1\mp\frac{E_{ij}}{|E_{ij}|}\right)+O\left(J^{2}\right)\co\nn
		z_{ij}^{\pm}&=\pm\ii|E_{ij}|-\frac{(D+1)J}{2D}+O\left(J^{2}\right)\ed
	\end{align}
	so we have
		\begin{align}
			(\mathcal{U}_1)_{i\not=j}&\sim e^{-\frac{(D+1)J}{2D}t}e^{-\ii E_{ij}t}\delta_{ii'}\delta_{jj'}+\frac{1}{D}\left(1-e^{-\frac{J}{2}t}\right)\delta_{ij}\delta_{i'j'}
			 \nn
			 &\nl-\frac{Je^{-\frac{(D+1)J}{2D}t}}{2D|E_{ij}|}\sin\left(|E_{ij}|t\right)\delta_{ij'}\delta_{ji'}\co
		\end{align}
	and 
	\begin{align}
		(\mathcal{U}_1)_{ii}&\sim e^{-\frac{(D+1)J}{2D}t}\cosh\left(\frac{J}{2D}t\right)\delta_{ii'}\delta_{ij'}+\frac{1}{D}\left(1-e^{-\frac{J}{2}t}\right)\delta_{i'j'}\nn
		&\nl+e^{-\frac{(D+1)J}{2D}t}\sinh\left(\frac{J}{2D}t\right)\delta_{ij'}\delta_{ii'}\ed
	\end{align}
	
	\paragraph{Large \( J \) expansion}
	\begin{align}\label{eq:largeJexp}
		g_{ij}&=1+O(J^{-2})\co\nn
		c_{ij}^{\pm}&=1\mp\frac{2\ii DE_{ij}}{J}+O\left(J^{-3}\right)\co\nn
		z_{ij}^{\pm}&=-\frac{(D+1)J}{2D}\pm\frac{J}{2D}+O\left(J^{-1}\right)\ed
	\end{align}
	so we have
	\begin{widetext}
		\begin{align}
			(\mathcal{U}_1)_{ij}&\sim e^{-\frac{D+1}{2D}Jt}\left[\cosh\left(\frac{Jt}{2D}\right)-\left(\frac{2\ii DE_{ij}}{J}\right)\sinh\left(\frac{Jt}{2D}\right)\right]\delta_{ii'}\delta_{jj'}
			+\frac{1}{D}\left(1-e^{-\frac{J}{2}t}\right)\delta_{ij}\delta_{i'j'}+e^{-\frac{D+1}{2D}Jt}\sinh\left(\frac{Jt}{2D}\right)\delta_{ij'}\delta_{ji'}\ed
		\end{align}
	\end{widetext}
    \bibliography{refer}
    \end{document}